\documentstyle[prd,aps,psfig]{revtex}
\bibstyle{unsrt}
\begin{document}
 
\preprint{LAUR-98-1914, HD-THEP-98-19}  
\twocolumn[\hsize\textwidth\columnwidth\hsize\csname 
@twocolumnfalse\endcsname
\draft
 
\title{Time evolution of correlation functions for
classical and quantum anharmonic oscillators}
\author{Lu\'{\i}s M. A.  Bettencourt$^1$ 
and Christof Wetterich$^2$}
\address{$^1$T6/T11, Theoretical Division, MS B288, 
Los Alamos National Laboratory, Los Alamos NM 87545} 
\address{$^2$Institut f\"ur Theoretische Physik, Universit\"at 
Heidelberg, Philosophenweg 16, 69120 Heidelberg, Germany}
\date{\today}
\maketitle
\begin{abstract}
The time evolution of the correlation functions
of an ensemble of anharmonic N-component  oscillators
with O(N) symmetry is described by a flow equation, exact
up to corrections of order $1/N^2$. We find effective irreversibility. 
Nevertheless, analytical and numerical
investigation reveals that the system does not reach thermal
equilibrium for large times, even when $N\rightarrow \infty$.
Depending on the initial distribution, the 
dynamics is asymptotically stable or it 
exhibits growing modes which break the conditions for 
the validity of the $1/N$ expansion for large time. 
We investigate both classical and quantum systems, the latter being 
the limit of an $O(N)$ symmetric scalar
quantum field theory in zero spatial dimensions.
\end{abstract}

\pacs{PACS Numbers : 11.15.Pg, 11.30.Qc, 05.70.Ln, 98.80.Cq 
\hfill    LAUR-98-1914,  HD-THEP-98-19} 

\vskip2pc]

Recently much progress was made in  understanding the 
evolution of quantum field theories away from thermal 
equilibrium in the leading $1/N$ approximation \cite{LORev}.
However this approach has its shortcomings. 
The omission of scattering leads to infinitely many conserved 
quantities \cite{Us} which prevent thermalization.

In \cite{Us} we have taken the first step in an investigation of the 
evolution of an $O(N)$ symmetric scalar field theory in next to leading 
order $1/N$.  
The resulting system describes the evolution of the 
4-point couplings in addition to the 2-point functions of the fields and 
their conjugate momenta. 
At this level we can hope to answer the  fundamental 
question of  whether the inclusion of scattering leads
to thermalization.

A full treatment of the non-linear behavior for the field theory 
requires much greater computational resources than at 
the leading order. One should therefore 
gain as much insight as possible in simple 
realizations of the problem \cite{0D}.
In this letter we study the limit of zero spatial dimensions of an 
$O(N)$ symmetric $\varphi^4$-theory, or, equivalently the evolution
of an ensemble of quantum anharmonic oscillators. 
 Our approach is based on the 
effective action which generates the equal time one particle irreducible (1PI)
correlation functions \cite{Wetterich,QWett}.

The time evolution of the effective action $\Gamma$ obeys
$\partial_t \Gamma[\phi, \pi;t] = - \left({\cal L_{ \rm cl}} 
+ {\cal L_{\rm q}} \right) \Gamma[\phi, \pi;t]$,
where the first operator, ${\cal L_{\rm cl}}$, generates the 
classical evolution, whereas ${\cal L_{\rm q}}$ 
determines the dynamics of the quantum effects.   
We will deal with the specific example of 
a $O(N)$ symmetric ensemble of anharmonic oscillators 
$Q_a$, $a=1,...,N$, obeying the microscopic equations of motion 
$\partial_t^2 Q_a = - m^2 Q_a - {\lambda\over 2} Q_b 
Q_b Q_a. $ 
The arguments of $\Gamma$ are the mean values of the coordinates 
$\phi_a = \langle Q_a \rangle$ and their conjugate momenta $\pi_a = 
\langle \dot Q_a \rangle $ (in the presence of sources).
From \cite{Wetterich,QWett}, one infers ($\psi_i \equiv (\phi_a,\pi_a)$)
\begin{eqnarray}
&& {\cal L}_{\rm cl} =   \pi_a 
{\delta \over \delta \phi_a} - m^2\phi_a
{\delta \over \delta \pi_a}
 - {\lambda \over 2} 
\left[\phi_b \phi_b \phi_a +  \phi_a 
G^{\phi \phi}_{bb}  \right. \nonumber 
\\ && \left.  + 2 \phi_b G^{\phi \phi}_{ba} -   
G^{\phi \psi}_{ai}
G^{\phi \psi}_{bj} G^{\phi \psi}_{bk} {\delta^3 \Gamma \over 
\delta \psi_i \delta \psi_j \delta \psi_k} \right] 
{\delta \over \delta \pi_a } ,  \label{e2} \\
&& {\cal L}_{\rm q} 
= {\lambda \over 8} \hbar^2~ \phi_a
{\delta \Gamma \over \delta \pi_b} 
{\delta \Gamma \over \delta \pi_b} 
{\delta  \over \delta \pi_a},
\label{e3}
\end{eqnarray}
On the level of 1PI 2 and 4 point functions
the most general form of $\Gamma$ consistent with $O(N)$ 
symmetry is 
\begin{eqnarray}
&& \Gamma = {1 \over 2}  
\Bigl\{  A \phi_a^*\phi_a  +   B  \pi_a^* \pi_a +  2 C \pi_a^*
\phi_a  \Bigr\} + {1\over 8} \Bigl\{ 
u  \phi_a \phi_a \phi_b \phi_b  \Bigr. \nonumber \\ \Bigl. &&
+ v \pi_a \phi_a \phi_b 
\phi_b +w \pi_a \pi_a \phi_b 
\phi_b  + s  \left[\pi_a  \pi_b  \phi_a  
\phi_b  - \pi_a \pi_a \phi_b 
\phi_b  \right]  \Bigr. \nonumber \\ \Bigl. && + y \pi_a \pi_a \pi_b 
\phi_b +  z 
\pi_a \pi_a \pi_b  \pi_b \Bigr\}  \label{e4} 
\end{eqnarray}
The time dependence of $A(t),\ldots,z(t)$ gives directly the time 
dependence of the propagator and the 4-point vertices. Higher vertices 
only induce corrections $\sim 1/N^2$. We emphasize that the truncation 
(\ref{e4}) still retains connected n-point functions of arbitrary 
order to the extent that they are constructed from 4-vertices and 
propagators ($G_{ab}^{\phi\phi} = G \delta_{ab} = 
\langle Q_a(t) Q_b(t) \rangle - \langle Q_a(t)\rangle \langle  
Q_b(t) \rangle$, {\it etc}.).
   
The exact equations for the inverse two point functions, with 
$G=B/(AB -C^2)$ and $c= C/B$, read
\begin{eqnarray}
&&\partial_t G = -2 c G,  \quad \partial_t B = 
- 2 \left( c + \gamma \right)  B,
 \label{e5} \\
&& \partial_t c = \omega^2 - {1 \over B G} 
+ \gamma c + c^2. \nonumber
\end{eqnarray}
Here, the time dependent frequency, $\omega$, and $\gamma$ are:
\begin{eqnarray}
&& \omega^2 = m^2 + {N+2 \over 2} \lambda  G  
\Bigl[ 1 - {G^2 \over 4} \left(4 u - 3 v c + 2 w  c^2  - y c^3 
\right)\Bigr], \nonumber  \\
&& \gamma = { N+2 \over 8} \lambda   G^3  \Bigl[  v  - 2 w c +3 y c^2  
-4 z  c^3   \Bigr].
\label{e6}
\end{eqnarray} 

For large $N$, with $\lambda, u,v,w,s,y,z$ scaling like $1/N$,
one may expand in small $1/N$. We see that 
$\gamma \sim 1/N$ and $\omega$ becomes independent of 
$u,v,\ldots,$ in leading order. In the limit $N \rightarrow \infty$
the system of evolution equations for the inverse 2-point functions
is closed. In this approximation $B/G$ becomes an additional
conserved quantity \cite{Us}. 

For a solution of (\ref{e5},\ref{e6}) 
at finite $N$, however, one needs the time
dependent 4-point functions. 
Their evolution  involves, in turn the 6-point functions. The exact system
is not closed and for any practical purpose we have to proceed to some 
approximation. Below we simply omit all contributions from 
1PI  6-point vertices.  Then 
the evolution equations for the quartic couplings are
\begin{eqnarray}
&&\partial_t u =   \omega^2 v + 4 \lambda B   c - \lambda \hbar^2 B^3 c^3 
-\lambda B   G^2 \Bigl\{ 2 (N+8) u c \Bigr. \nonumber \\
&&  \Bigl. \qquad - (N+8) v c^2 
+ \bigl[ (N+2) w - (N-1)s \bigr] c^3   
\Bigr\},  \nonumber \\
&& \partial_t v =   2 \omega^2  w  
- 4 u - \gamma v + 4 \lambda B  - 3 \lambda \hbar^2 B^3  c^2
-  \lambda B G^2 \Bigl\{    \Bigr. \nonumber \\ 
&& \Bigl. \  2 (N+8) u + \bigl[ (N-10) w -  3 (N-1)s \bigr] c^2  
+(N+8) y c^3   
\Bigr\},     \nonumber \\
&&\partial_t w =  3 \omega^2 y - 3v - 2 \gamma w 
 - 3 \lambda \hbar^2 B^3  c -  \lambda B G^2   
\Bigl\{  (N+8) v \Bigr.  \nonumber \\
&&   \Bigl. \qquad + \bigl[ (N-10) w - 3(N-1) s \bigr] c
+2  (N+8) z c^3 \Bigr\}, 
\nonumber  \\
&&\partial_t s =  2 \omega^2 y - 2 v  - 2 \gamma s
- 2 \lambda  \hbar^2 B^3 c -\lambda  B G^2 \Bigl\{ (N + 6) v \Bigr.
\nonumber \\
&& \Bigl.  \qquad - 2 \bigl[ 4 w + (N-1) s \bigr] c 
+ (N+2) y c^2   +8 z c^3    \Bigr\}, \label{e7} \\ 
&& \partial_t y =  4 \omega^2 z - 2w - 3 \gamma y  
- \lambda \hbar^2 B^3 -   \lambda B G^2  
\Bigl\{ (N+2) w  \Bigr. \nonumber \\
&&   \Bigl.    \qquad -  (N-1) s 
- (N+8) y c  + 2 (N+8) z c^2 \Bigr\}, \nonumber \\
&&\partial_t z =  - y  - 4 \gamma  z. \nonumber 
\end{eqnarray}
Here we have omitted some terms $\sim 1/N^2$. 
The neglected 6-point functions can also be treated 
consistently as being $\sim 1/N^2$.
Thus, our set of evolution Eqs.~(\ref{e5}-\ref{e7}) contains 
all contributions in the next to leading order $1/N$. 
The average energy $ \epsilon = \epsilon_1 + \epsilon_2 
\equiv \langle H \rangle $,
\begin{eqnarray}
&& \epsilon_1 = {N \over 2} \Biggl\{
B^{-1} + \Bigl[ m^2 + c^2 + {N+2 \over 4} \lambda  G \Bigr] G \Biggr\},  \\
&& \epsilon_2 = - {N \left( N+2 \right)\over 8 } 
\lambda  G^4
\Bigl[ u - v c  + w c^2 - y  c^3 + z c^4 \Bigr], 
\nonumber
\end{eqnarray} 
is conserved by the exact evolution equations, as well as by our 
truncation. Our system meets at least some basic conditions 
for thermalization: The number of interacting degrees of 
freedom $N$ can be arbitrarily large and scattering 
effects are included.

A first hint that the approach to equilibrium for large times
is not obvious, however, comes from the study of possible time 
invariant solutions. 
The fixed points of the system Eqs.~(\ref{e5}-\ref{e7}) correspond
to stationary probability distributions. They 
obey, for $\omega^2> 0$,
\begin{eqnarray}
&& G^{-1}_* = \omega_*^2 B_*, \quad   c_* = 0, \quad v_*=0, 
\quad  y_*=0,  \nonumber \\ 
&& u_* =    {\lambda  B_* 
+  \left (  \omega_*^2/2 \right) w_* \over 1 
+ {N+8 \over 2} \lambda/(\omega_*^4 B_*) }, \label{e8} \\
&& w_* =   {2  \omega_*^2 z_* +  {N-1 \over 2} \lambda  
\left( \omega_*^4 B_* \right)^{-1} 
s_* - {\lambda \over 2} \hbar^2   B_*^3  
\over  1+ {N+2 \over 2} \lambda/(\omega_*^4 B_*) }. \nonumber
\end{eqnarray}
We observe a large manifold of fixed points since Eqs.~(\ref{e8}) have 
solutions for arbitrary $B_*$, $s_*$ and $z_*$ !
This property seems not to be an artifact of the truncation.  
The  fixed point manifold 
becomes even larger if one includes the six point functions and seems 
to characterize any higher (finite) polynomial truncation.
Classical thermodynamic equilibrium, at temperature $T$ 
(for $\hbar=0$) corresponds to the 
particular point in this manifold:
\begin{eqnarray}
&& B_{\rm eq} = \beta =1/T, \quad  c_{\rm eq}= 0, \quad 
G_{\rm eq}=  {T \over \omega_{\rm eq}^2} \nonumber \\
&& v_{\rm eq} = w_{\rm eq} = s_{\rm eq} = y_{\rm eq} = z_{\rm eq} = 0, 
\label{e9}  \\
&& u_{\rm eq} = {\lambda \over T}  \left( 1 + {N+8 \over 2} 
{\lambda T \over \omega_{\rm eq}^4}  \right)^{-1}.
\nonumber
\end{eqnarray}
where $\omega_{\rm eq}^2$ obeys 
\begin{eqnarray}
&& \left( \omega_{\rm eq}^4 
- m^2 \omega_{\rm eq}^2 - {N+2 \over 2} \lambda T \right) 
\left( 1 + {2 \over N+8} {\omega_{\rm eq}^4 \over \lambda T} \right) 
\nonumber \\  
&& = - { N+2 \over N+8 } \lambda T,
\label{e10}
\end{eqnarray}
with ${\rm lim}_{ N \rightarrow \infty} \ \omega_{\rm eq}^2 = 
{1 \over 2} \left(m^2 + \sqrt{ m^4 + 2 N \lambda T} \right).$  

The thermal fixed point in the quantum system occurs for 
non-zero $w_*$,$z_*$. 
In particular, we find in the quantum system a fixed point which differs 
from the classical equilibrium fixed point only by 
$z_* = \lambda \hbar^2  B_*^3/(4 \omega_*^2)$, 
instead of $z_{\rm eq}=0$. 
For this fixed point, all correlation functions involving only 
$\phi$ are the same as for classical equilibrium. Inversely, for every 
fixed point of the quantum system there is one for the 
classical system, differing only by a shift in $z_*$. Up to this shift, the 
quantum equilibrium could be reached by the classical system!
    
An obstruction to thermalization arises from the existence of 
an infinite set of conserved quantities.
Besides $\langle H \rangle$, also arbitrary powers 
$\langle H^p (L^2)^q \rangle$ (with $L^2$ the squared generalized angular 
momentum) are conserved. For thermal equilibrium $\langle L^2 \rangle$ 
or $\langle H^2 \rangle - \langle H \rangle^2$ are computable
as functions of $T$, eg. $\langle L^2 \rangle_{\rm eq}= N(N-1) 
T^2/\omega_{\rm eq}^2$.
Initial values of conserved quantities 
away from equilibrium prevent thermalization.

For classical statistics ($\hbar=0$), we have shown \cite{Us} 
that the dynamical system describing the linearized 
motion orthogonal to the fixed point manifold  
is characterized by three doubly degenerate eigenvalues which are 
strictly imaginary. Consequently the motion is purely oscillatory.
To probe the behavior in
the non-linear regime, away from the fixed point manifold,
we have solved the system of differential Eqs.~(\ref{e5}-\ref{e7}) 
numerically for various initial conditions.
Unless otherwise stated, we consider below the the parameters
$N=10$, $m^2=1$, $\lambda = 1/N$ and $\hbar=0$.  

\begin{figure}
\centerline{\psfig{file=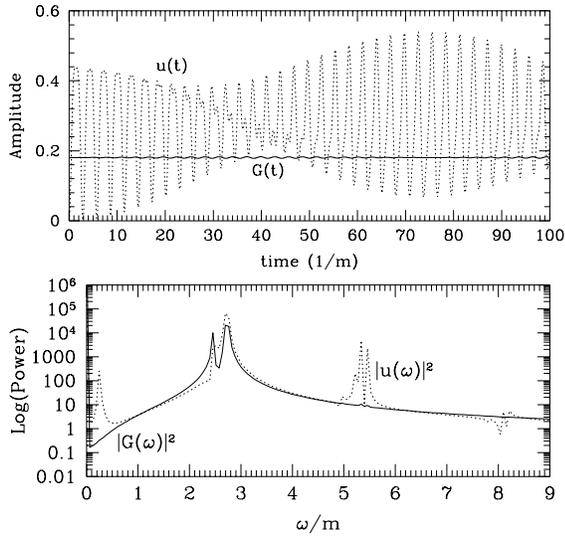,width=3.0in}}
\caption{$G(t)$ and $u(t)$ for Gaussian initial 
conditions and $\hbar=0$. The second plot shows the Fourier 
power spectrum.} 
\label{fig1}
\end{figure}

\begin{figure}
\centerline{\psfig{file=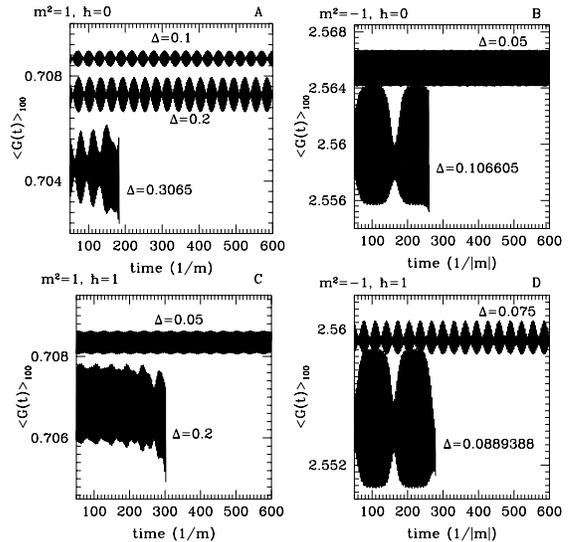,width=3.0in}}
\caption{$\langle G(t)\rangle_{100}$ for initial 
conditions of (\ref{e11}) with $T=1$.} \label{fig2}
\end{figure}

We first show, in Fig.~\ref{fig1},  
the result of Gaussian initial conditions
with $G(0)=0.18069$, $B(0)=5$ and  all other initial couplings zero.
After a rapid initial increase of $u$ 
the typical behavior becomes purely 
oscillatory. This confirms the results of our small 
fluctuation analysis and the Fourier power spectrum reflects the 
relevant frequencies. On the other hand, the system never comes back 
to the initial Gaussian distribution which would require all quartic 
couplings to vanish simultaneously. 
Due to nonlinearities it clearly exhibits effective irreversibility!
The oscillatory behavior
indicates  that the thermal values for the correlation functions 
can never be approached in a strict sense. Nevertheless, the time averages 
of some correlation functions over sufficiently large time intervals $\tau$
could still  approach the thermal
equilibrium values. We therefore investigate quantities of the type
$\langle G \rangle_\tau (t) = (\tau)^{-1} 
\int_{t-\tau/2}^{t+\tau/2} dt'  G(t')$. Typical 
initial conditions for our study of the approach to equilibrium
are
\begin{eqnarray}
&& G(0) =  G_{\rm eq}(T)(1 +\Delta), \quad u(0) = u_{\rm eq}(T), 
\nonumber \\
&& c(0) = v(0) = y(0) =w(0)=s(0) = z(0)=0, \label{e11} \\
&& B(0)^{-1} = T - \Delta G_{\rm eq}(T) \left[ m^2 
 + {N+2 \over 4} \lambda G_{\rm eq}(T) (2 + \Delta) \right. \nonumber \\ 
&& \left. - {N+2 \over 4} \lambda G^3_{\rm eq}(T) u_{\rm eq}(T) 
 \left( 4 + 6 \Delta + 4 \Delta^2 + \Delta^3 \right) \right] \nonumber \\
&& + {N+2 \over 4} \lambda G^4(0) \left(u(0) -u_{\rm eq}(T) \right).
\nonumber
\end{eqnarray}
Here $T$ is the equilibrium temperature and $G_{\rm eq}(T)$, $u_{\rm eq}(T)$
the corresponding values in Eqs.~(\ref{e9}). 
The initial condition for $B(0)$ is chosen such that the energy of the system 
$\epsilon = {N \over 4} \left( 3 T + G_{\rm eq} (T) m^2 \right),$
is independent of $\Delta$. Thus, the initial conditions correspond 
all to the same effective temperature and the large time behavior 
should become independent of $\Delta$ if the system thermalizes.  
In Fig.~\ref{fig2} we plot $\langle G \rangle_{100} (t)$  for the 
initial conditions of (\ref{e11}).
We see that for $\Delta $ not too large $\langle G \rangle_{100}$ 
has for large $t$ a stable mean value independent 
of $\tau$. The mean value depends, however, 
on the initial condition as specified by $\Delta$. The 
curves correspond to the same effective $T$, but to different 
initial $\langle L^2 \rangle$. For $\Delta >0$ we find 
a clear signal that the '2-point function' 
$\langle Q^2 \rangle$ never thermalizes. This 
constitutes the most important result of this letter.

The situation for the quantum system is not very different, as seen 
by comparing Figs.~\ref{fig2}A and \ref{fig2}C. 
For negative $m^2=-1$ we show in Figs.~\ref{fig2}B and \ref{fig2}D
the results of the classical ($\hbar=0$) and quantum ($\hbar=1$) 
evolution for $T=1$ and with $G_{\rm eq}(T),\ u_{\rm eq}(T)$ 
adapted appropriately. 

We repeated the analysis of Fig.~\ref{fig2} 
for different $N$ with similar results,  even for 
very large $N$. The dependence of 
$\langle G \rangle$ on $\Delta$ becomes independent of $N$ as  
$N\rightarrow \infty$.    

\begin{figure}
\centerline{\psfig{file=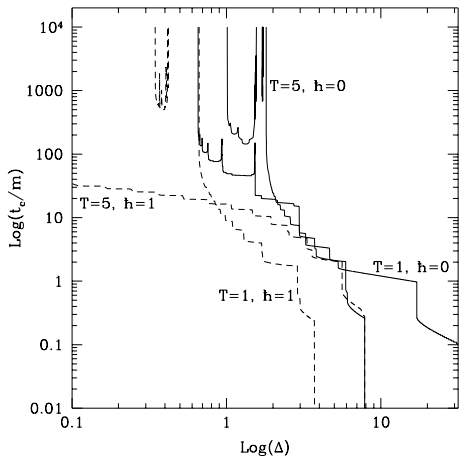,width=2.5in}}
\caption{$t_c$ vs. $ \Delta $, for $T=1,5$ and $B(0)=1,0.2$. 
The discontinuities seem to arise at arbitrary small steps in $\Delta$. 
For $ \Delta $ such that no point is plotted $t_c>10^4$.}
\label{fig3}
\end{figure}

The odd number of degrees of freedom and the absence of more  
conservation laws hints to the fact that the full nonlinear system
of flow equations should not be Hamiltonian, albeit being energy 
conserving. The spectrum of 
a Hamiltonian system is semi-positive definite and the system 
is stable in the sense that it has no growing modes. Otherwise 
this property is lost. Of course the existence of 
one growing mode implies others since the energy must be preserved 
by the evolution. 
This is precisely what happens to the full nonlinear system in the 
strong anharmonic regime. 
As already visible in Fig.~\ref{fig2} the system becomes 
unstable if the initial
conditions are far from the fixed point manifold. More precisely this 
means that the couplings grow large and the validity of the $1/N$ 
expansion breaks down for large $t$. A useful criterion for the validity
of the $1/N$ expansion requires that the contribution to $\epsilon$
proportional to $u$, $\vert \epsilon_u \vert = {N (N+2) \over 8} \lambda 
\vert G^4 u \vert$ remains smaller then ${1 \over 3} \epsilon_1$. We denote
by $t_c$ the time when this condition is violated and plot $t_c$ 
as a function of $\Delta$ in Fig.\ref{fig3} for $T=1$ and $T=5$.
The initial conditions are chosen here without the $\Delta$ terms in $B(0)$,
i.e. $B(0)=1/T$. 
For $\Delta \leq \Delta_c$ the $1/N$ expansion remains valid at least for 
$t \leq 10^4$. (The critical values are $\Delta_c= 0.655(1.012)$ 
for $T=1(5)$ and $\hbar=0$
and $\Delta_c=0.346(\Delta_c \leq 10^{-6})$ for $T=1(5)$ and $\hbar =1$).  
For $\Delta  > \Delta_c$ one observes a very peculiar behavior of 
$t_c(\Delta)$ with discontinuities at certain values of $\Delta$.
It suggests a chaotic dependence  of the correlation
functions on the initial conditions! (This type of behavior has been 
observed at leading $1/N$ \cite{Chaos}). This should not be confounded
with the much more common chaotic behavior for individual solutions
of the equations of motion. The latter, once averaged in an 
ensemble, often leads to regular behavior of the 
correlation functions. Note that 
there are restrictions on the initial values of the correlation 
functions as they must be consistent with a positive probability
distribution. We have not systematically investigated these conditions
so far. 

\begin{figure}
\centerline{\psfig{file=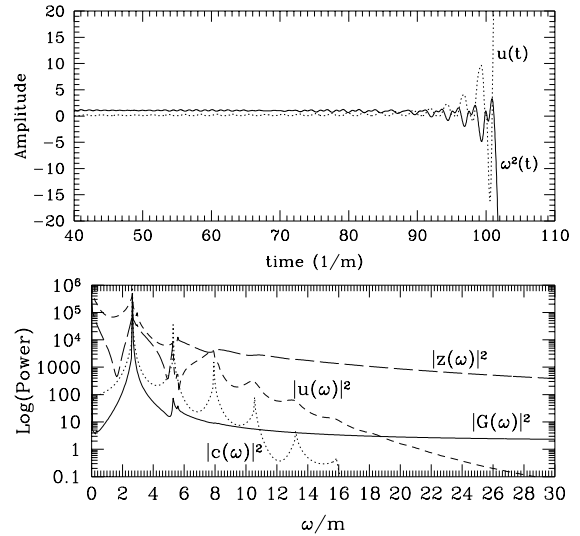,width=3.0in}}
\caption{The evolution of $\omega^2(t)$, $u(t)$  for initial conditions with 
$T=1$, $\Delta=0.666$ and $\hbar=1$ and corresponding 
the frequency power spectra
for $ G,c,u $ and $ z $.} 
\label{fig4}
\end{figure}

Below we address the behavior for $\Delta > \Delta_c$
in more detail in order to shed some light on the origin of the instability.
Fig.\ref{fig4} shows a typical evolution for initial conditions
generating instabilities. Notice that $u(t)$ grows unbounded 
for large times while $\omega^2(t)$  becomes negative. 
The power spectrum shows how the system goes stiff, particularly
for the 4-point functions. In order to ensure the accuracy of our
numerical integration we have used two independent methods, one based 
on an adaptative step algorithm and another especially tailored 
to handle stiff problems. 
Although the latter is intrinsically less precise we observed 
that instabilities occurred for coincident $t_c$. We  
conclude they are inherent to the dynamical system and 
robust to the choice of numerical integrator.  

From Fig.~\ref{fig4} it is  apparent that 
many frequencies are shared among the 2 and 4-point functions. 
This may lead to secular instabilities since products 
of the 2-point functions act as sources for the dynamical equations 
of the 4-point functions. Secular instabilities result from 
an oscillator  being driven at its own natural frequency 
and result initially in linear growth of its amplitude.  
For a given frequency $\omega$ shared by all degrees 
of freedom odd products of the 2-point functions can
generate secular growth of the 4-point functions. In the quantum case 
these terms are more common, which may explain 
why the quantum motion generally becomes unstable sooner.

In conclusion, we have presented a detailed study of the time
evolution of correlation functions for ensembles of $N$-component quantum
and classical anharmonic oscillators. Our truncation method converges
for large $N$ and is well suited to describe an approach to thermal 
equilibrium. Nevertheless, we find that the isolated systems do not 
thermalize!  It will be very interesting to see if similar features
persist in field theory or if effective thermalization occurs 
in the presence of many frequencies for the linear 
problem, due to dephasing  \cite{LORev,Us}.

We thank F.~Cooper,  S.~Habib and T.~Papenbrock for useful suggestions.   
This work was supported in part by the {\it Deutsche 
Forschungsgemeinshaft} and the European Science Foundation.

\end{document}